\documentclass[aps,prd,preprint,tightenlines,superscriptaddress,nofootinbib]{revtex4}
\usepackage{epsfig}
\usepackage{graphicx}
\usepackage{psfrag}
\usepackage{amsmath,amssymb}
\usepackage{colordvi}
\usepackage{amsfonts}
\usepackage{enumerate}
\usepackage{slashed}
\usepackage{color}
\usepackage{xcolor}

\begin{document}

\title{Ionization Yield from Nuclear Recoils \\ in Liquid-Xenon Dark Matter Detection}

\author{Wei Mu}
\affiliation{INPAC, Department of Physics, and Shanghai Key Lab for Particle Physics and Cosmology, Shanghai Jiao Tong University, Shanghai, 200240, P. R. China}
\author{Xiangdong Ji}
\affiliation{INPAC, Department of Physics, and Shanghai Key Lab for Particle Physics and Cosmology, Shanghai Jiao Tong University, Shanghai, 200240, P. R. China}
\affiliation{Maryland Center for Fundamental Physics, University of Maryland, College Park, Maryland 20742, USA}
\date{\today}
\vspace{0.5in}

\begin{abstract}
The ionization yield in the two-phase liquid xenon dark-matter detector has been studied in keV nuclear-recoil energy region. The newly-obtained nuclear quenching as well as the recently-measured average energy required to produce an electron-ion pair are used to calculate the total electric charges produced. To estimate the fraction of the electron charges collected, the Thomas-Imel model is generalized to describing the field dependence for nuclear recoils in liquid xenon. With free parameters fitted to experiment measured 56.5~keV nuclear recoils, the energy dependence of ionization yield for nuclear recoils is predicted, which increases with the decreasing of the recoiling energy and reaches the maximum value at 2$\sim$3~keV. This prediction agrees well with existing data and may help to lower the energy detection threshold for nuclear recoils to $\sim$1~keV.
\end{abstract}

\maketitle

\section{Introduction}
In recent years, liquid xenon (LXe) has emerged as a leading medium for detecting weakly interacting massive particles~(WIMPs), which is one of the most interesting dark matter~(DM)~candidates~\cite{Aprile:2010bt,Aprile:2011dd,Akerib:2012ys,Abe:2013tc,Li:2012zs}. Many advanced LXe detectors~\cite{Aprile:2006pr,Sorensen:2010,Manzur:2010} work in the two-phase mode with measurements of both scintillation (or direct scintillation light, denoted as S1) and ionization (proportional scintillation light, denoted as S2) signals at the same time. The scintillation and ionization signals, together with the relative scintillation efficiency (${\cal L}_{\rm eff}$) and the ionization yield (${\cal Q}_{\rm y}$) respectively, have been used to reconstruct the nuclear recoil energy. Thus ${\cal L}_{\rm eff}$ and ${\cal Q}_{\rm y}$ are crucial physical quantities for the energy calibration of WIMPs detection.

According to a recent theoretical calculation in Ref.~\cite{Publishing}, ${\cal L}_{\rm eff}$ drops rapidly with decreasing energy in the low energy region, especially below 3~keV important for light DM particle detection ($\sim$ 10 GeV/c$^2$). In this case, the uncertainty in ${\cal L}_{\rm eff}$ becomes an important factor for reconstruction of the nuclear recoil energy. Because the collection of primary scintillation photons becomes more difficult at low recoiling energy, it is hard to measure scintillation efficiency experimentally. Therefore, the energy threshold of the LXe detectors determined from the primary scintillation efficiency is relatively high. On the other hand, the ionization signal will become more suitable for several reasons: First, the ionization and scintillation signals in LXe are anti-correlated~\cite{Aprile:2007,Sorensen:2010jc}. The anti-correlation refers to the phenomenon that the ionized electrons escape recombination with positive ones ($Xe_2^+$), thereby reducing the scintillation photons. Therefore, in low-energy region, the ionization signal can be more prominent than scintillation. Second, single electrons are measured with higher efficiency than single photons~\cite{Sorensen:2010jc}. In addition, when a strong external electric field, $E_d$, is applied across the LXe detector, a larger fraction of ionized electrons may escape from the event site and become ionization signals. Therefore, the ionization signals are a better choice to detect in low recoil energy region and will help lower the energy thresholds of LXe detectors. However, although a series experimental measurements and theoretical studies have been done for the relative scintillation efficiency (${\cal L}_{\rm eff}$), only few measurements are available for the ionization yield (${\cal Q}_{\rm y}$)~\cite{Hitachi:2005,Aprile:2006pr,Sorensen:2010,Manzur:2010,Sorensen:2010jc,Bezrukov:2011}.

Serving as a link between the ionization signal and the deposited energy of the WIMPs in LXe, the ionization yield is defined as the number of observed electrons per unit recoil energy (e$^-$/keV):
\begin{equation}
{\cal Q}_{\rm y} (E_{\rm nr})=\frac{Q(E_{\rm nr})}{E_{\rm nr}} \ ,
\end{equation}
where $Q(E_{\rm nr})$ is the number of electric charges collected from nuclear recoil with initial energy $E_{\rm nr}$. This value is expected to depend on the external electric field but be independent of any specifics of a xenon detector.

In this paper, we present a theoretical study of the energy dependence of ionization yield for nuclear recoils. To obtain ${\cal Q}_{\rm y}$ theoretically, one has to calculate the actual collectable charges, $Q$. The latter in turn depends on the following two quantities: 1) the total charges produced for nuclear recoils, $Q_0$, and 2) the ionization efficiency, $r=Q/Q_0$. In the present study, we use the nuclear quenching factor, also called Lindhard factor $q_{\rm nc}(E_{\rm nr})=\eta(E_{\rm nr})/E_{\rm nr}$, calculated in Ref.~\cite{Publishing} and the newly measured average energy to produce an electron-ion pair, also called $W$-value, in Ref.~\cite{Seguinot:1992} to calculate the number of total charges produced by the nuclear recoils, $Q_0=E_{\rm nr} q_{\rm nc}(E_{\rm nr})/W$.  We use the generalized Thomas-Imel model to predict the field and energy dependence of the ionization efficiency, $r=Q/Q_0$, taking into account the effects of the ionization density and track structure of nuclear recoils. Consequently we find the ionization yield,
\begin{equation}
\label{eq:Final}
{\cal Q}_{\rm y} (E_{\rm nr})= \frac {E_{\rm nr} q_{\rm nc}(E_{\rm nr})}{W} \times \frac{Q}{Q_0}(E_{\rm nr}) \times \frac{1}{E_{\rm nr}} = \frac {q_{\rm nc}(E_{\rm nr})}{W} \times \frac{Q}{Q_0}(E_{\rm nr}) \ .
\end{equation}
which will be compared with available experimental data.

This paper is organized as follows. In Section 2, we review the ionization process of nuclear recoils and the measured $W$-value in different experiments. Combining theoretically calculated Lindhard factor with experimentally measured $W$-value, we calculate the number of total charges generated by nuclear recoils. In Section 3, we review the electron-ion recombination process and existing theoretical models for recombination rate. Then we generalize the Thomas-Imel recombination model to parameterize the field dependence of ionization yield for nuclear recoils by considering the specific track structure of nuclear recoils. In Section 4, we present the model prediction for energy dependence of the ionization yield from nuclear recoils. Finally, we summarize and discuss our results in Section 5.

\section{Electronic Energy Dissipation and Total Charge Yield}

In this section, we review the recent theoretical results on the total electronic energy dissipation (Lindhard factor) and the experimentally-measured average energy required to produce electron-ion pair, hence calculate the total charge yield from nuclear recoils.

\subsection{Electronic Energy Dissipation and Nuclear Quenching}

When a xenon nucleus is elastically scattered by a DM particle, it becomes a recoiling xenon atom (or nucleus) with kinetic energy up to a few tens keV \cite{Lewin:1996ap}. During the slowing down process, the recoiling xenon atom will excite and ionize other atoms inside the LXe medium, and at the same time produce secondary recoils, which subsequently lead to a collision cascade together with secondary trajectories surrounding the principle trajectory of the initial recoiling atom. After all the recoiling xenon atoms in the collision cascade are thermalized, the initial kinetic energy of the recoiling xenon atom dissipates into the kinetic energy of xenon atoms, excitation and ionization energies of atomic electrons. The electronic excitations will give rise to photons (scintillation signals) and free electrons (ionization signals), respectively.

\begin{figure}[hbt]
\begin{center}
\includegraphics[width=12cm]{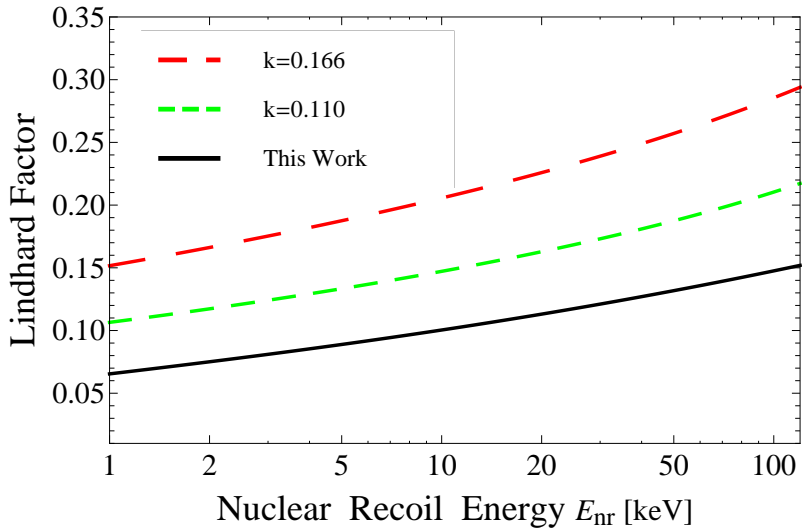}
\caption{The Lindhard factor, $q_{\rm nc}(E_{\rm nr})=\eta (E_{\rm nr})/E_{\rm nr}$, of LXe shows the fraction of energy given to electrons.
The red dashed curve is the empirical result from Eq.~(\ref{eq:L}) setting k=0.166, which is widely quoted by different authors while the green dashed curve is the result when setting k=0.110, which is calculated by Hitachi. The black solid curve is used in this work and obtained by using the EPS fitted to the available experimental data.}
\label{fig:Lfactor}
\end{center}
\end{figure}

Thus, to calculate the charge yield, one has to know the total electronic energy dissipation of the recoiling atom. The whole collision cascade process is in principle quantum mechanical and involves complicated many-body physics. One normally uses the Lindhard factor, also called nuclear quenching factor, to estimate the fraction of the energy given to electrons. This quantity was originally derived from Lindhard's basic integral equation in Ref.~\cite{Lindhard:1963}, which is an approximation for the collision cascade,
\begin{equation}
\label{eq:L}
q_{\rm nc}(E_{\rm nr})=\frac{\eta (E_{\rm nr})}{E_{\rm nr}}=\frac{k g(\epsilon)}{1+k g(\epsilon)} \ ,
\end{equation}
where $\eta (E_{\rm nr})$ is the energy transferred to electrons when the initial nuclear recoil energy is $E_{\rm nr}$. $g(\epsilon)$ is an empirical expression which can be found in Ref.~\cite{Lewin:1996ap} and $k$ is the proportionality constant between the electronic stopping power $(dE/dx)_{\rm el}$ and the velocity of the projectile (recoil atom). For xenon, Lindhard proposed a value $k=0.166$~\cite{Lindhard:1963}. Considering the collision cascade, Hitachi re-calculated the electronic stopping power of recoiling xenon atoms in a LXe target and derives a smaller value $k=0.110$~\cite{Hitachi:2005}. However, both authors overestimate the total electronic energy dissipation due to ignoring the faster fall-off behavior of the electronic stopping power (ESP) in low energy region. Based on the ESP measured in low energy region (40$\sim$200~keV)~\cite{Fukuda:1981}, we have performed a calculation of Lindhard factor in Ref.~\cite{Publishing} in which the result is much lower than that given by Lindhard and Hitachi. In Fig.~{\ref{fig:Lfactor}}, we show the results. From experimental data, the nuclear recoils produce more scintillation light than the electronic ones, and hence the Lindhard quenching factor shall be smaller than relative scintillation efficiency ${\cal L}_{\rm eff}$. Clearly our result satisfy this requirement.  Thus here we adopt the new result for the Lindhard quenching.

\subsection{Average Energy Required to Produce One Electron-Ion pair}
To describe the average energy required to produce one electron-ion pair, Platzman's phenomenological theory, which is originally proposed for high-energy radiations in rare gases in Ref.~\cite{Platzman:1961}, is widely referenced as a theoretical explanation. Platzman assumes the total dissipated electronic energy can be divided into three parts: ionization, excitation and sub-excitation electrons:
\begin{equation}
\label{eq:EnergyBlance}
E^{ee}= N_i E_i + N_{ex} E_{ex} + N_i \epsilon \ ,
\end{equation}
where $E^{ee}$ is the electronic energy dissipation, $N_i$ is the number of electron-ion pairs produced at an average energy expenditure of $E_i$, and $N_{ex}$ is the number of excitons at an average energy expenditure of $E_{ex}$. $\epsilon$ is the average kinetic energy of sub-excitation electrons. From Eq.~{\ref{eq:EnergyBlance}}, one can derive the expression for $W$-value,
\begin{equation}
\label{eq:Wvalue}
W= E^{ee}/N_i= E_i + (N_{ex}/N_i) E_{ex} + \epsilon \ .
\end{equation}
The Platzman's original paper gave a rough estimations for $E_i$, $N_{ex}/N_i$ and $\epsilon$, which generates the $W$-value qualitatively.

Experimentally, $W$-value for ionization have been measured in LXe by several groups ~\cite{Robinson:1973,Konno:1973,Takahashi:1974,Takahashi:1975,Huang:1977,Aprile:1991,Seguinot:1992} over the past forty years. The authors use $\alpha$ sources, $\gamma$-rays, X-rays or mono-energetic electrons sources respectively and get quite different $W$-values for LXe, which are listed in Table~{\ref{tab:W-value}}. It is not surprising that the literature values of $W$ are scattered in these measurements since absolute charge measurements are inherently challenging. Firstly, it is very difficult to identify the exact energy dissipation of the radiations precisely. Due to this reason, high linear energy transfer (LET) and low-energy radiations will be the good sources. Secondly, the electron-ion recombination exists, even for large drift electric fields, therefore it is hard to identify what fraction of generated electrons are collected. Because of that, one has to choose some low LET radiations as the sources and apply recombination theory to estimate the recombination rate, where the ``experimental result'' would depend also on which recombination theory is used.

\begin{table}[htbp]
\centering
\begin{tabular}{l l l l}
\hline
Radiation~Source & ~~W-value~(eV)~~ & ~~Year~~ & Reference\\
\hline
X-Ray & ~~7.3 & ~~1973 & Robinson and Freeman~\cite{Robinson:1973}\\
$^{210}$Po & ~~17.3 & ~~1973 & Kinno and Kobayashi~\cite{Konno:1973}\\
$^{63}$Ni & ~~16.5$\pm$1.4 & ~~1974 & Takahashi {\it et al}~\cite{Takahashi:1974}\\
$^{207}$Bi & ~~15.6$\pm$0.3  & ~~1975 & Takahashi {\it et al}~\cite{Takahashi:1975}\\
X-Ray & ~~15.2 & ~~1977 & Huang and Freeman~\cite{Huang:1977}\\
$^{207}$Bi & ~~15.1$\pm$1.79 & ~~1991 & Aprile {\it et al}~\cite{Aprile:1991}\\
Electron~beams& ~~9.76$\pm$0.70 & ~~1992 & Seguinot {\it et al}~\cite{Seguinot:1992}\\
\hline
\end{tabular}
\caption{The W-value in liquid xenon from different experiment measurements}
\label{tab:W-value}
\end{table}

Based on above discussion, we believe the low-energy electron beams are the better radiation sources to measure the $W$-value and we examine two individual experimental measurements from Takahashi et al. ~\cite{Takahashi:1975} and Seguinot et al.~\cite{Seguinot:1992}, both of which use electron sources. In 1975, Takahashi {\it et al.} use an internal-conversion electron source (1~MeV) to produce electron pulses in LXe and compare the signal with a 5.5~MeV $\alpha$ particles signal in an argon-methane (5\%) gas mixture, whose $W$-value is known~\cite{Takahashi:1975}. Using the relative pulse heights for the two materials, the authors obtain the $W$-value in LXe as $W=15.6{\pm}0.3~\rm {eV}$, where the ratio of the $W$-value to the band gap energy, $E_g$, of LXe (9.3~eV) is 1.68. In 1992, Seguinot {\it et al.} use mono-energetic electron beams ($\sim$100~keV) through a 12 $\mu$m thick foil of Mylar to ionize and excite LXe~\cite{Seguinot:1992}. In this experiment, MeV's to GeV's energies are deposited in a LXe test cell. In the meantime, the authors also performed measurements using both fixed energy deposition with variable applied electric fields as well as fixed applied electric fields with variable energy deposition to check the impact from electron-ion recombination. Then the authors apply Jaffe-like formula and Thomas-Imel formula to estimate the recombination of electron-ion pairs. The $W$-value they get is much lower than previous measurements where $W=1.05E_g=9.76{\pm}0.70~\rm {eV}$, which is just slightly higher than the band gap energy of LXe. Seguinot's result has been disputed in Ref.~\cite{Miyajima:1995,Seguinot:1995}. However, Seguinot {\it et al.} claim that their measurement is accurate for low-energy electron excitation of LXe.

Takahashi {\it et al.} uses Platzman's phenomenological theory to explain their results: set $E_i=1.13E_g$ implicitly and $N_{ex}/N_i=0.06$ theoretically; calculate the $\epsilon$ using Shockley's model as $\epsilon=0.48E_g$. Then they get the theoretical prediction for $W$-value as $1.66E_g$ which is consistent with their experimental value ($1.68E_g$). Seguinot {\it et al.} did not give a theoretical explanation for their results in their publishing. However, it appears that if one ignores the sub-excitation electrons in Eq.~(\ref{eq:EnergyBlance}), and then set $E_i=E_g$ and $N_{ex}/N_i=0.06$ (theoretical value in Ref.~\cite{Takahashi:1975}) or $N_{ex}/N_i=0.20$ (experimental value in Ref.~\cite{Doke:2002}), we can get a theoretical prediction for W-value between $1.05E_g$ and $1.18E_g$ which agrees reasonable well with Seguinot's experimental value ($1.05E_g$).

Although, Takahashi's measurement is widely quoted, we suspect that Seguinot's measurement may be more reasonable for low energy electron recoils and nuclear recoils in the WIMPs energy range. The kinetic energy for the sub-excitation electrons, $\epsilon$ in Eq.~(\ref{eq:Wvalue}), comes from the binary collision between an incident particle and a free (ionized) electron of the media. For low energy electron recoils or nuclear recoils, few electronic energy dissipated as kinetic energy of sub-excitation electrons compared with that from high-energy particle recoils, because the energy transferred to the free (ionized) electrons through binary collision between the incident nucleus (or low energy electron) and the ionized electrons is negligible ($\sim$m$_{\rm el}$/m$_{\rm ncl}$) (or small). Therefore, the electronic energy dissipations may mainly produce electron ionization or excitation. Additionally, the sub-excitation electrons are unobservable through experiment while $E_g$, $E_{ex}$ and $N_{ex}/N_i$ are well measured or calculated and identified by different experiments or theories. Therefore, we choose to adopt Seguinot's measurement for $W$-value 9.76~eV in the following sections. 

\section{Recombination of Electron-Ion Pairs and Ionization Yield}

In this section, we use the phenomenological model to study the electron-ion recombination process and from which, determine the ionization yield.

\subsection{Recombination Process and Thomas-Imel Model for Electron Recoil}

After knowing the total electronic energy dissipations and the $W$-value, we can easily calculate the total electric charge produced by the radiation. However, the charge collected in the detector is not necessarily equal to that initially produced by the radiation. The main reason is the electron-ion's recombination, which greatly reduces the final ionization. Electrons released from xenon atoms in ionization events undergo a random motion under the influence of mutual electrostatic interactions. Part of the oppositely charged particles (electrons and ions) approach the others to a short distance and finally recombine which results in reduction of the ionization yield. In the presence of an external electric field, more of the pairs are permanently separated and more electrons are expected to escape from recombination as the electric field increases. Electron-ion recombination exists at any finite field strength. Therefore, the correct interpretation of data obtained from LXe detectors requires good understanding of the recombination processes in this medium.

The physics of recombination in LXe is not well understood yet. Phenomenologically three kinds of recombination may occur in LXe detector:
\begin{itemize}
\item Initial recombination, which refers to the process where an electron, freed from an atom and thus producing a positive ion, returns to its origin and recombines to produce a neutral atom again. Under no external electric field scenario, this mechanism is presumed to be the leading of the recombination of most pairs;
\item Bulk recombination, which occurs when there are a continuum of charges of both electrons and ions presenting at the same location. A random electron combines with a random positive ion at a rate proportional to the densities of the electron and ion;
\item Columnar recombination, which occurs when an energetic charged particle produces a column of ion pairs in an external electric field. As the electrons drift in one direction and the positive ions in the other, electrons and ions occasionally recombine.
\end{itemize}

From the phenomenological consideration of electron-ion recombination, a large number of theoretical studies~\cite{Langevin:1903,Onsager:1938,Jaffe:1940,Kramers:1952,Thomas:1987pa,Mozumder:1995,Chabod:2009} have been made for over a century. Among these, Thomas and Imel gave a concise analytical expression in Ref.~\cite{Thomas:1987pa} after considering the recombination behavior for electron recoils or alpha in LXe.

Thomas and Imel start from the diffusion equations for positive ion $(n^+)$ and electrons ($n^-$) proposed by Jaffe in Ref.~\cite{Jaffe:1940}:
\begin{eqnarray}
\partial_t n_i^+ &=& -\mu_+ \vec{E_d}\cdot\nabla n_i^+ + d_+ \nabla^2 n_i^+ -\alpha n_i^- n_i^+\ ,   \nonumber \\
\partial_t n_i^- &=& -\mu_- \vec{E_d}\cdot\nabla n_i^- + d_- \nabla^2 n_i^- -\alpha  n_i^+ n_i^- \ , 
\end{eqnarray}
where $\mu_{\pm}$ is the mean mobilities of the ions and electrons under external electric field, while $d_{\pm}$ and $\alpha$ are coefficients corresponding to the diffusion and recombination terms. To simplify the equations, they ignored the diffusion terms, considered to be much smaller than the electron drift, and also dropped the drift terms of the positive ion. Solving the equations in the box boundary conditions, Thomas and Imel obtained the following result:
\begin{eqnarray}
\notag
\frac{Q}{Q_0} =\frac{N_i^{\rm esc}}{N_i} = \frac{1}{\xi}\ln(1+\xi); ~~~~   \xi = \frac{n_i \alpha}{4a^2 \mu_- E_d} \ , 
\end{eqnarray}
where $\alpha/(4a^2 \mu_-)$ is a constant determined by the dielectric constant, the electron mobility and the ionization volume length scale. The diffusion processes are neglected in case of a large external electric field because the diffusion speeds of electric charges are negligible compared to their drift speeds in the electric field. Thus, to get more reliable results at smaller electric field, we have to consider the electron mobility terms $\mu_- E_d$ in $\xi$.

In LXe, when electrons are excited to the conduction band from the valence band by an energetic radiation, they become free electrons and drift in external electric field. At low fields, the electron drift velocity, $v_d$, is almost proportional to the field strength, $E_d$, with the electron mobility, $\mu$, as the proportionality constant ($v_d=\mu E_d$). At high fields, about 10~kV/cm in LXe~\cite{Miller:1968}, the electron drift velocity saturates and becomes independent of electric field strength. However, at low field, we shall consider the random fluctuations of electric fields acting on the free electrons. This electric fields cause the free electrons to diffuse out of the positive ion region with a statistically constant speed $v_0$, which means some electrons may escape from recombination even when no external field is applied. Additionally, using the electronic stopping power or LET as a rough estimation of the charge density, we can generalize $\xi$ as:
\begin{equation}
\label{eq:xi}
\xi=\frac{n_i \alpha}{4a^2 (\beta v_0+ v_d) }\equiv \frac{K_1}{1+K_2\times E_d} \left(\frac{dE}{dx}\right)_{\rm el} \ .
\end{equation}
where $\beta$ is some unknown constant.

In Ref.~\cite{Publishing}, $K_1$ is considered as a field-independent parameter and fitted to experimentally measured LET dependence of the scintillation yields in LXe~\cite{Doke:1988nb}, from which $K_1=2.53$~MeV$^{-1} \cdot$g$\cdot$cm$^{-2}$. In this work, we fit $K_2$ to the experimental data for electron recoils and alphas in Ref.~\cite{Aprile:2006pr}, and find $K_2=3$~KV$^{-1}\cdot$cm. Here we do not try to re-calculate $Q/Q_0$ in Ref.~\cite{Aprile:2006pr} for the ionization yield when fitting the parameter $K_2$ since we just want to check if the generalized Thomas-Imel model can obtain the recombination trend accurately.

\begin{figure}[hbt]
\begin{center}
\includegraphics[width=12cm]{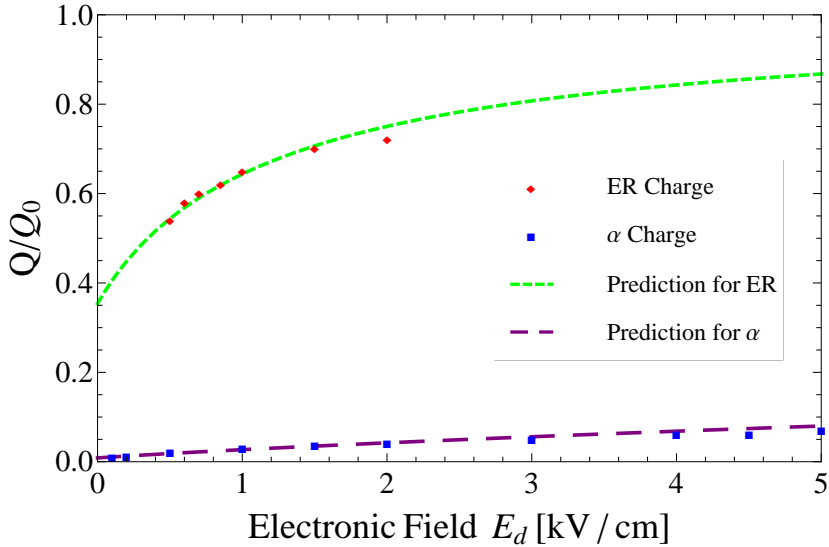}
\caption{Reproduction of field dependence of ionization yield efficiency in LXe for 122~keV electron recoils (ER) and 5.5~MeV alphas.
The data is extracted from Fig.~3 in Ref.~\cite{Aprile:2006pr}.}
\label{fig:RproE}
\end{center}
\end{figure}

Both electron recoils, which have lower LET value and suffer less recombination, and alphas, which have higher LET value and suffer more recombination, have clear field dependence shown Fig.~{\ref{fig:RproE}}. Thomas-Imel box model successfully reproduces the field dependence. It yields the recombination probability as a function of the ionization density and the external electric field, where a smaller number of escaping electrons are produced along a denser track (higher LET), at the same time, the number of escaping electrons increases with the increasing of the applied electric field. 

\subsection{Generalization of Thomas-Imel Model to Nuclear Recoils}

When it comes to nuclear recoils, the electric field dependence is quite different. The ionization density dependence of recombination still exists, however no significant field dependence of recombination has been measured~\cite{Aprile:2006pr}. There must be some unidentified physics effect, except for the different LET values, which may interfere the recombination behavior of nuclear recoils and has not been covered in original Thomas-Imel box model. Chepel {\it et al.} analyzed the track structures for electronic recoils and alphas~\cite{Chepel:2012}. The tracks of electron recoils (with low LET) can be regarded as a line of widely populated positive ions with average distance close to the Onsager radius. The free electrons reach thermal energy at sufficiently large distances from the track escape from recombination, causing the increase of ionization yield for electron recoils in external electric field. For alphas (high LET), the track structure is a little different, which can be described as concentric cylinders consisting of a central core and a surrounding penumbra~\cite{Hitachi:2005}. A continuous line of positive charges forms the core of the main trajectory. Compared with electron recoils, the charge density is much higher. However, both electron recoils and alphas have a line-like or column-like track structure, though the charge density along the trajectory is different. 

For nuclear recoils, it is a totally different physical picture. Although the track structure of nuclear recoils is not fully understand, it is clear that nuclear recoils will cause a large number of secondary nuclear recoils through binary collisions during the slowing down process, which results in a tree-like track structure, where lots of secondary trajectories are surrounding the main trajectory. This behavior has been proved by computer simulations~\cite{Dahl:2009pt,Ziegler:2010}. The secondary trajectories enlarge the spatial distribution of the ions projecting to the main trajectory. Although the electric charges in the secondary trajectory is sparse, they screen the applied electric field significantly, which in some sense is similar to the electrostatic screening effect. It is reasonable to surmise that the screen effect can explain why the recombination for nuclear recoils depends weakly on external electric field. The significantly-varied physics pictures for different radiation particles hinder the development of a universal theory of recombination in LXe detector. Existing recombination models mainly consider the column-like or line-like radiation track structures, ignoring the specific track structure of nuclear recoils. It is not surprising that existing recombination models cannot quantitatively explain the field dependence of recombination for nuclear recoils. To get more solid and reliable prediction, we have to extend the existing models.

Here, we generalize Thomas-Imel box model starting from the box boundary condition. The box boundary condition assumes both electrons and ions are distributed in a certain volume and model their evolution due to the charge drift and reactions. This model well covers the cylindrically symmetric distributions of electrons and ions along the trajectory of electron recoils and alphas, which have column-like structures. But for nuclear recoils, the tree-like track structure breaks the field dependence relation. The neglecting of the geometry of the secondary trajectories distribution for nuclear recoils causes the model fail to predict the recombination rate for nuclear recoils. Based on this, we can generalize Thomas-Imel box model by introducing a field-screen degree $S$ in $\xi$ to correct the electric field dependence:
\begin{equation}
\label{eq:xi2}
\xi=\frac{K_1}{1+K_2\times E_d^{1/S}} \left(\frac{dE}{dx}\right)_{\rm el} \,
\end{equation}
where the screen exponent $S$ is assumed to be proportional to the spatial distribution scale of the track structure. Since the Bohr impulse principle, normally used to assess the radiation core radius, is not applicable for nuclear recoils, we have to find another physical quantity to evaluate the scale of the nuclear recoils' track in this article. As we know, the secondary trajectories are caused by binary nuclear recoils, where nuclear stopping power (the energy loss to nuclei per unit path $(dE/dx)_{\rm ncl}$), is a useful physical quantity to describe the nuclear collision. So we may relate $S$ to $(dE/dx)_{\rm ncl}$ through $S=K_3\times (dE/dx)_{\rm ncl}$. Fitting to experimental data, we can identify the free parameters $K_2$ and $K_3$.

\begin{figure}[hbt]
\begin{center}
\includegraphics[width=12cm]{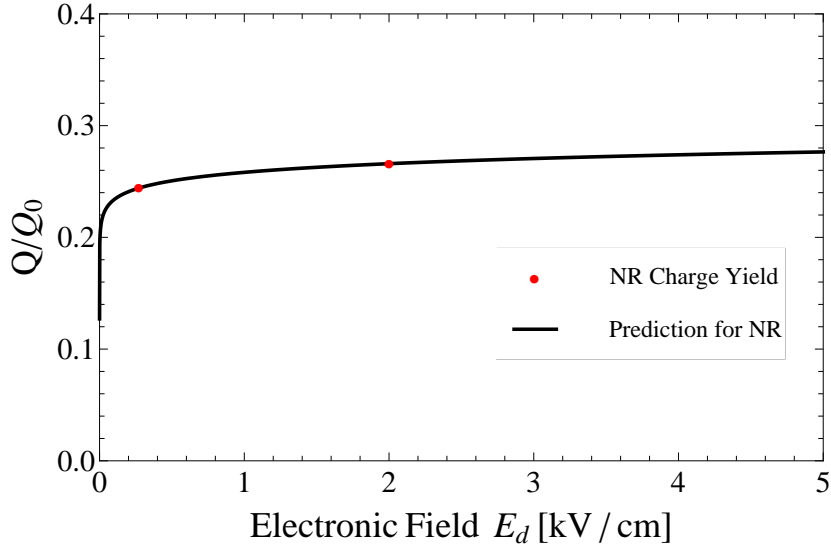}
\caption{Reproduction of electric field dependence of the ionization yield efficiency in LXe from 56.5~keV nuclear recoils (NR).
The charges collected is from Fig.~4 in Ref.~\cite{Aprile:2006pr}, the number of the total charges produced is calculated
from $Q_0=E_{\rm nr} q_{\rm nc}(E_{\rm nr})/W$, where $L$ is from our recent work, Ref.~\cite{Publishing}, and $W=9.76$~eV is measured by Seguinot {\it et al.} in Ref.~\cite{Seguinot:1992}. The nuclear stopping power $(dE/dx)_{\rm ncl}$ used is from Ref.~\cite{Ziegler:2010}.}
\label{fig:RproN}
\end{center}
\end{figure}

\begin{figure}[hbt]
\begin{center}
\includegraphics[width=12cm]{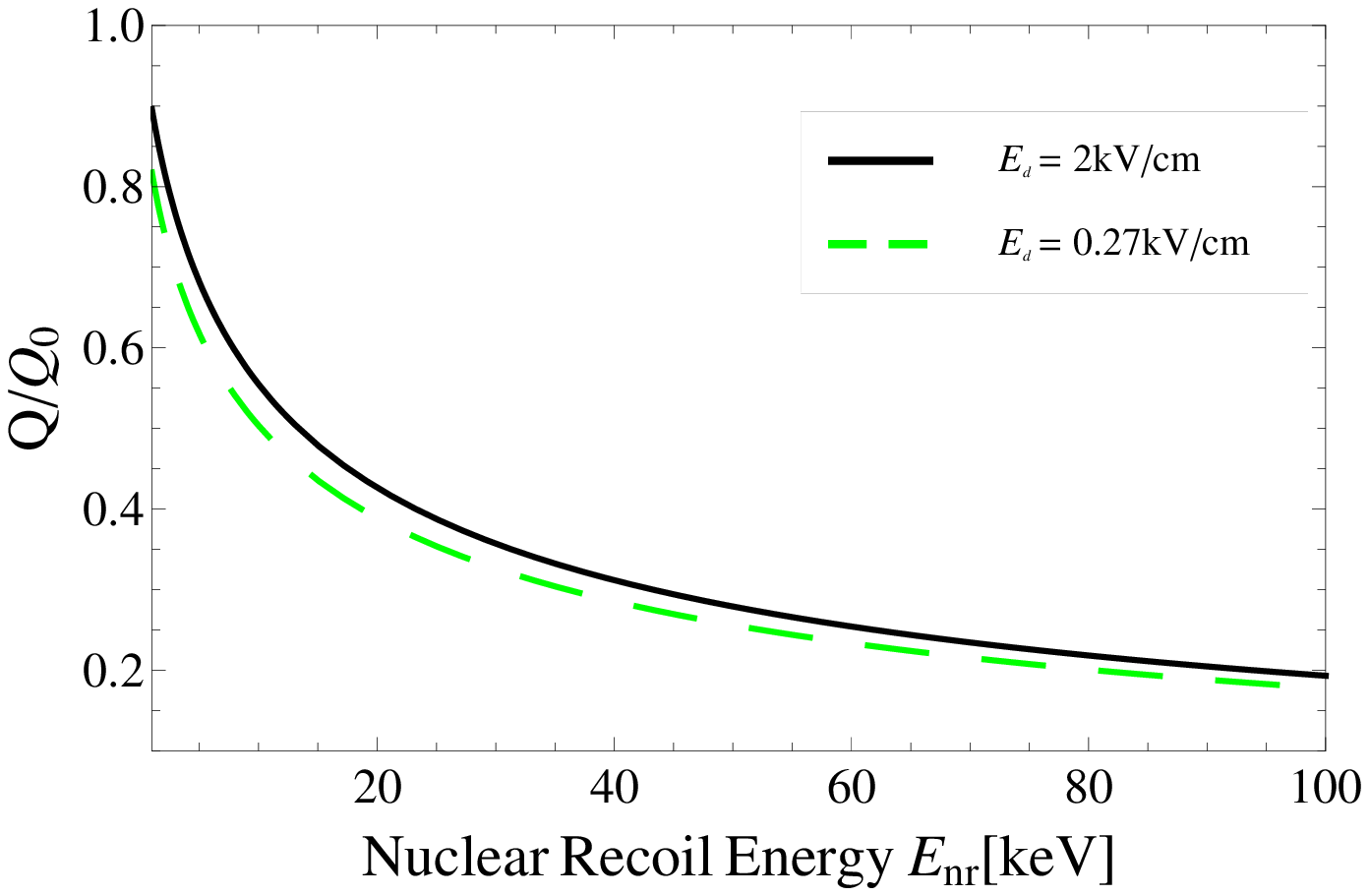}
\caption{Energy dependence of the ionization yield efficiency, $Q/Q_0$, in LXe under external electric fields 2~kV/cm (solid black curve) and 0.27~kV/cm (green dashed curve) for nuclear recoils.
We can see the ionization yield for nuclear recoils in LXe depends on electric field weakly.}
\label{fig:RecRate}
\end{center}
\end{figure}

Our goal is to quantitatively describe the electric field dependence of the recombination rate for nuclear recoils already observed, and then predict the energy dependence of the ionization yield for nuclear recoils in LXe. Therefore, we first fit $K_2$ and $K_3$ to the experimentally measured electric-field dependence of the ionization yield in LXe for 56.5~keV nuclear recoils~\cite{Aprile:2006pr}. The charges collected is from Fig.~4 in Ref.~\cite{Aprile:2006pr}, the number of the total charges produced is re-calculated in our work from $Q_0=E_{\rm nr} q_{\rm nc}(E_{\rm nr}) /W$, where $q_{\rm nc}$ is from our recent work, Ref.~\cite{Publishing}, and $W=9.76$~eV is measured by Seguinot {\it et al.} in Ref.~\cite{Seguinot:1992}. The nuclear stopping power $(dE/dx)_{\rm ncl}$ used is from Ref.~\cite{Ziegler:2010}. We obtain from the fit that $K_2=14.755$ when the unit for $E_d$ is kV/cm, and $K_3=0.004$~MeV$^{-1} \cdot$g$\cdot$cm$^{-2}$. The experimental data and the fit curve from the generalized Thomas-Imel model are shown in Fig.~{\ref{fig:RproN}}.

Using the fitted free parameters, we can predict the energy dependence of the ionization for nuclear recoils. The results for two electric fields, 2~kV/cm and 0.27~kV/cm are shown in Fig.~{\ref{fig:RecRate}}. As the recoil energy gets lower, the density of the ions also decreases, the recombination effect becomes less important. In this case, the ionization efficiency increase. We see this trend clearly starting from 40 keV. It reaches 100\% at zero recoil energy. Again, the electric field dependence of the ionization efficiency is small in the nuclear recoils. 

\section{Prediction for Energy Dependence and Comparison with Data}

\begin{figure}[hbt]
\begin{center}
\includegraphics[width=16cm]{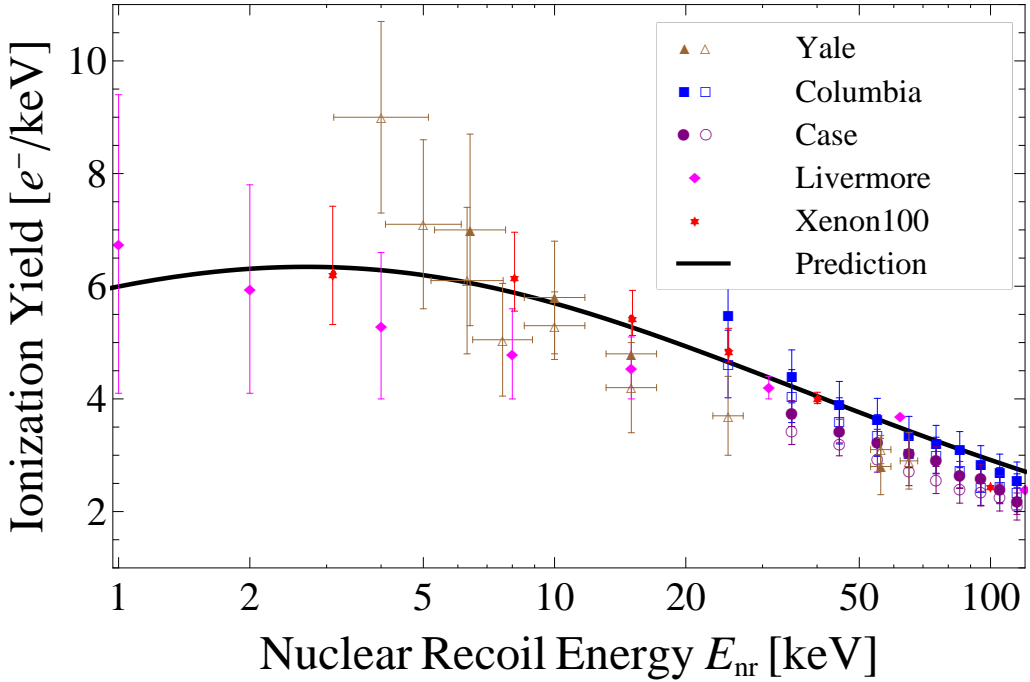}
\caption{Energy dependence of the ionization yield, ${\cal Q}_{\rm y}$, obtained from the generalized Thomas-Imel Model, compares with the available experimental data.
Here, we use the Lindhard factor in Ref.~\cite{Publishing} to evaluate the fraction of energy given to electrons and $W$-value 9.76~eV measured by Seguinot {\it et al.} in Ref.~\cite{Seguinot:1992} to calculate the total charges produced. }
\label{fig:CompE}
\end{center}
\end{figure}

Combining the total electronic energy dissipation in Ref.~\cite{Publishing}, $W$-value in Ref.~\cite{Seguinot:1992} with the generalized Thomas-Imel model for electron-ions recombination in previous section, we can predict the ionization yield for nuclear recoils in LXe detector using Eq.~(\ref{eq:Final}). The  energy dependence of ionization yields ${\cal Q}_{\rm y}$ are shown as the black solid curve in Fig.~{\ref{fig:CompE}}, in the presence of a 2~kV/cm external electric field.

Along with our prediction, we have also shown the experimental data from different measurements: The brown triangles are from Yale group~\cite{Manzur:2010}; The blue squares are the result from the Columbia group measurement~\cite{Aprile:2006pr}, from which our free parameters are fitted to the 56.5~keV data; The purple circles are from the Case group~\cite{Aprile:2006pr}; The magenta diamonds are from \cite{Sorensen:2010} and the red six-point-stars are from the XENON100 collaboration with mono-energy neutron and broad spectrum neutron sources~\cite{Aprile:2013teh}. Our prediction is in good agreement with the most recent data within experimental uncertainties.

The important feature in our result is that the ionization yield increases with decreasing energy, and reaches a maximum value at the recoiling energy in 2$\sim$3~keV region. This maximum is produced by the increasing ionization efficiency and decreasing electronic energy dissipation at low recoil energy. It will be very interesting to test this prediction experimentally. Since the scintillation efficiency rapidly decreases with decreasing energy, particularly when less than 3~keV, the anti-correlation behavior of ionization yield helps to lower the low-energy threshold of LXe detectors.

\section{Conclusion}

In this paper, we have studied the ionization yield of nuclear recoils in LXe which is very important for DM searches. Specifically we combine the theoretically calculated Lindhard factor with the experimentally measured $W$-value to calculate the total charge generated for nuclear recoils. By analyzing available experimental data and the difference among electron recoils, alphas, and nuclear recoils, including the ionization density, the track structure and so on, we generalize Thomas-Imel model to the nuclear recoils and reproduce the field dependence of ionization yield for 56.5~keV nuclear 
recoils in LXe. Then, we use this model to predict the energy dependence of ionization yield under fixed external electric field in LXe, and the
the result satisfies the existing experimental data well. The result may help to lower the low-energy threshold for two-phase 
LXe detectors.


\section{Acknowledgment}
W. M. thanks Fei Gao for helpful comments and discussions. This work is partially supported by a 973 project, No. 2010CB833005, of China's Ministry of Science and Technology, and a grant (No. 11DZ2260700) from the Office of Science and Technology in Shanghai Municipal Government. This work has also been supported by the U.S. Department of Energy via grant DE-AC02-05CH11231,

\end{document}